\documentclass[12pt]{article}
\makeatletter
\@addtoreset{equation}{section}
\makeatother

\usepackage{amssymb,mathrsfs,multirow,color}

\def\Eqn#1{Eq.\ (\ref{#1})}
\def\Eqs#1#2{Eqs.\ (\ref{#1}) and (\ref{#2})}
\def\3Eqs#1#2#3{Eqs.\ (\ref{#1}), (\ref{#2}) and (\ref{#3})}

\def\vec#1{\mathchoice
{\mbox{\boldmath $#1$}}
{\mbox{\boldmath $#1$}}
{\mbox{\boldmath $\scriptstyle #1$}}
{\mbox{\boldmath $\scriptscriptstyle #1$}}}

\def\mag#1{{\tt #1}}
\def\ket#1{|#1\rangle}
\let\tilde=\widetilde
\let\hat=\widehat
\let\bar=\overline
\def\ms#1{\mathscr{#1}}
\def\next{\nonumber\\}
\def\Wrep#1{\breve{#1}}

\def\bmatrix#1{\left[ \matrix{#1} \right]}

\title{\bf Dirac, Majorana and Weyl fermions}

\author{\bf Palash B. Pal\\
Saha Institute of Nuclear Physics\\ 
1/AF Bidhan-Nagar, Calcutta 700064, India}

\date{}

\begin{document}


\maketitle

\begin{abstract}

  This is a pedagogical article which discusses various kinds of
  fermion fields: Dirac, Majorana and Weyl.  The definitions and
  motivations for introducing each kind of fields is discussed, along
  with the connections between them.  It is pointed out that these
  definitions have to do with the proper Lorentz group, and not with
  respect to any discrete symmetry.  The action of discrete symmetries
  like charge conjugation and CP on various types of fermion fields,
  particularly important for Majorana fermions, has also been
  clarified.

\end{abstract}

\section{Introduction}
When Dirac first wrote down his relativistic equation for a fermion
field, he had primarily electrons in mind.  It doesn't require much
mind-reading to deduce this conclusion, because his first article
\cite{Dirac:1928hu} on this issue was entitled ``The Quantum Theory of
the Electron''.  Electrons have mass and charge.  In his solutions,
Dirac found the antiparticle, which has the same mass as the electron
but is different from the electron because it has opposite charge.

Dirac's paper was published in 1928.  The very next year, Weyl
\cite{Weyl:1929fm} showed that for massless fermions, a simpler equation
would suffice, involving two-component fields as opposed to the
four-component field that Dirac had obtained.

And then, in 1930, Pauli \cite{Pauli:1930pc} proposed the neutrinos to
explain the continuous energy spectrum of electrons coming out in beta
decay.  The neutrinos had to be uncharged because of conservation of
electric charge, and they seemed to have vanishing mass from the
analysis of beta decay data.  It was therefore conjectured that the
neutrinos are massless.  Naturally, it was assumed that the neutrinos
are therefore Weyl fermions, i.e., their properties are described by
Weyl's theory.

There was also the possibility that neutrinos are the antiparticles of
themselves, since they are uncharged.  Description of such fermion
fields was pioneered by Majorana \cite{Majorana:1937vz} in 1937.  The
question was not taken seriously because, at that time, everybody was
convinced that neutrinos are Weyl fermions.

The question became important much later, beginning in the 1960s, when
people started examining the consequences of small but non-zero
neutrino masses, and possibilities of detecting them.  If neutrinos
have mass, they cannot be Weyl fermions.  This opened the discussion
of whether the neutrinos are Dirac fermions or Majorana fermions.

Majorana fermions became important in Particle Physics for other
reasons as well.  Supersymmetric theories require Majorana fermions as
partners of spin-0 or spin-1 bosonic fields.  One might also add that
supersymmetric theories are best described in superspace, obtained by
augmenting the usual spacetime variables with some fermionic
parameters which transform as Majorana spinors.

With all the experience of working with Dirac fermions, working with
Majorana fermions produced some hiccups.  Even now, it is not uncommon
to see fantastic claims about Majorana particles or fields in the
literature that come out of strange jugglery or gymnastics with these
objects: complicated operations that often have no rational or
analytical basis~\cite{Singh:2006ad}.

The uneasiness can be compared to a feeling that Alice had experienced
during her travels in Wonderland.  At one point, she drank something
and became very small.  Then she saw a small cake with the words `EAT
ME' marked on it.  She wondered whether she would shrink further, or
grow back to her original size if she ate it.  Finally, she
\begin{quote}
ate a little bit, and said anxiously to herself, `Which way? Which
way?', holding her hand on the top of her head to feel which way it
was growing, and she was quite surprised to find that she remained the
same size...
\end{quote}
After this, Lewis Carroll comments that in fact, there was nothing to
be surprised about.  ``This generally happens when one eats
cake''.  But Alice, by that time, got so much accustomed to seeing the
extraordinary that she was getting surprised by seeing an ordinary
thing happening to her.

Majorana fermions are quite simple objects, simpler than Dirac
fermions.  But we are so much accustomed to Dirac particles that we
try to understand Majorana particles through Dirac particles.  This is
a roundabout way, and creates problems.  In this article, we will
introduce Majorana fermions through an imaginary journey in which we
will pretend that we do not know about Dirac fermions.  Dirac fermions
will be also be mentioned, for the sake of completeness.  And,
although it is now known that the neutrinos are indeed massive, Weyl
fermions will appear in our journey as well.  Although no known (or
even conjectured) particle can possibly be a Weyl fermion, we will see
that the concept is very useful, because Weyl fermions can be seen as
building blocks of any fermion field.

Our journey will be anachronistic.  In the title of the paper, we have
listed the three kinds in alphabetical order.  That is also not the
order we will follow in the article.

\section{The Klein-Gordon equation and its solutions}
To begin this journey, let us not even worry about fermions.  In
relativistic physics, the Hamiltonian of a free particle of mass $m$
must satisfy the equation
\begin{eqnarray}
H^2 = \vec p^2 + m^2 \,,
\label{kg.dispersion}
\end{eqnarray}
written in the natural unit in which we have chosen $c=1$.  When we
try to build a quantum theory, we can let both sides act on the
wavefunction.  Using the standard co-ordinate space operator for the
momentum and setting $\hbar=1$ by the choice of units, we obtain the
equation for the wavefunction $\phi$ to be
\begin{eqnarray}
\left( {\partial^2 \over \partial t^2} - \vec \nabla^2  + m^2 \right)
\phi = 0 \,.
\label{kg.kg}
\end{eqnarray}
This is the Klein-Gordon equation.  

The differential operator acting on $\phi$ is real.  So, if we choose
an initial condition in which $\phi$ is real everywhere, the evolution
through the equation will keep it real.  This will give us a real
solution of the Klein-Gordon equation.

Plane waves of the form $e^{-ip^\mu x_\mu}$ are solutions to the
Klein-Gordon equation provided
\begin{eqnarray}
p^\mu p_\mu &=& m^2 \,.
\label{pp}
\end{eqnarray}
We can use them to expand any other solution.  For real solutions,
such a Fourier expansion will be:
\begin{eqnarray}
\phi(x) = \int_p \Big( a(p) e^{-ip\cdot x} + a^\star (p) e^{+ip\cdot x}
\Big) \,, 
\label{kg.phi}
\end{eqnarray}
We have divided the Fourier terms into two parts by imposing the
condition 
\begin{eqnarray}
p^0 &>& 0 \,,
\label{p0>0}
\end{eqnarray}
so that the reality condition is transparent.  The measure of the
integral over $p$ has been left undefined, and will be kept so,
because it is not important for our discussion.  For quantum fields,
the Fourier co-efficients $a(p)$ become operators, and $a^\star(p)$
should be understood to be the hermitian conjugate of $a(p)$.

\section{The Dirac equation and its solutions}\label{s:DE}
\subsection{The equation}
After this preamble, let us discuss the Dirac equation, which is
\begin{eqnarray}
\Big( i \gamma^\mu \partial_\mu - m \Big) \Psi = 0 \,.
\label{DiracEq}
\end{eqnarray}
The equation can be seen as the Schr\"odinger equation,
\begin{eqnarray}
i {\partial \Psi \over \partial t} = H \Psi \,,
\end{eqnarray}
arising from the Hamiltonian
\begin{eqnarray}
H = \gamma^0 \Big( \gamma^i p^i + m \Big) \,.
\label{DiracH}
\end{eqnarray}
Alternatively, it can be seen as the Euler-Lagrange equation coming
from the Lagrangian
\begin{eqnarray}
\ms L = \bar\Psi \Big( i \gamma^\mu \partial_\mu - m \Big)
\Psi \,, 
\label{Lag}
\end{eqnarray}
where $\bar\Psi\equiv\Psi^\dagger\gamma^0$.

In these equations, $\gamma^\mu$ denotes a collection of four
matrices, each $4\times4$, which satisfy the conditions
\begin{eqnarray}
\Big[ \gamma^\mu , \gamma^\nu \Big]_+ &=& 2 g^{\mu\nu} \,, 
\label{clifford}\\*
\gamma_0 \gamma_\mu \gamma_0 &=& \gamma_\mu^\dagger \,,
\label{gamdag}
\end{eqnarray}
where $[A,B]_+=AB+BA$ denotes the anticommutator.  The first one,
which has an implied unit matrix on the right hand side, is necessary
so that the Dirac equation complies with the energy-momentum relation
of \Eqn{kg.dispersion}.  The second equation, which is necessary so
that the Hamiltonian implied by the Dirac equation is hermitian, can
more explicitly be written as
\begin{eqnarray}
\gamma_0^\dagger = \gamma_0 \,, \qquad \gamma_i^\dagger = - \gamma_i
\,. 
\label{gamdag0i}
\end{eqnarray}

Let us set up some notational rules that will be helpful for avoiding
confusion.  Any solution of \Eqn{DiracEq} will be called a {\em
  fermion field} and will be denoted by $\Psi(x)$.  For specific
solutions, we will use different notations.  For example, for Majorana
fields, we will use the notation $\psi(x)$, whereas for Weyl fields,
$\chi(x)$.  The word {\em spinor} will be used to denote any
column-like function of energy and momentum which, when multiplied by
a factor $\exp(ip\cdot x)$ or $\exp(-ip\cdot x)$, becomes a solution
of the Dirac equation.

\subsection{Real solutions}
Is the Dirac equation a real equation like the Klein-Gordon equation?
The answer depends on what the $\gamma^\mu$'s are.  If all non-zero
elements of all four $\gamma^\mu$'s are purely imaginary, then
\Eqn{DiracEq} is real.  So the question is: can we define the
$\gamma^\mu$'s, subject to their basic properties encrypted in
\Eqs{clifford}{gamdag}, so that they are purely imaginary?

Indeed, we can.  This was first found by Majorana.  The
representation, denoted by a tilde on the
matrices,\footnote{Throughout, we use the notation that whenever an
  array will be enclosed in square brackets, each entry should be
  thought of as a block of length 2, i.e., a $2\times2$ matrix for
  square arrays, and a $2\times1$ column for a column array.}  is
this:
\begin{eqnarray}
\tilde\gamma^0 = \bmatrix{ 0 & \sigma^2 \cr \sigma^2 & 0 }
\quad &,& \quad 
\tilde\gamma^1 = \bmatrix{ i\sigma^1 & 0 \cr 0 & i\sigma^1 
} \,,\nonumber\\
\tilde\gamma^2 = \bmatrix{ 0 & \sigma^2 \cr -\sigma^2 & 0
} \quad &,& \quad 
\tilde\gamma^3 = \bmatrix{ i\sigma^3 & 0 \cr 0 & i\sigma^3 
} \,,
\label{majorep}
\end{eqnarray}
where the $\sigma^i$'s are the usual Pauli matrices, written such that
$\sigma^2$ is imaginary while the other two are real.  Clearly, 
\begin{eqnarray}
\tilde \gamma_\mu^\star = \null - \tilde \gamma_\mu 
\label{gamstar}
\end{eqnarray}
as proposed.  These matrices constitute the Majorana representation of
the $\gamma$-matrices.

So now suppose that we have written down the Dirac equation in detail,
using the matrices from \Eqn{majorep}.  That will be a real
equation, just like the Klein-Gordon equation.  Therefore, one should
be able to find real solutions to this equation.  In other words, we
will find solutions which satisfy
\begin{eqnarray}
\tilde \psi = \tilde \psi^\star \,.
\label{psi=psi*}
\end{eqnarray}
Such solutions will represent Majorana fermions.  We emphasize that
\Eqn{psi=psi*} is valid in the Majorana representation, a fact that is
remembered by the presence of the tilde on top.

And now, this is the problem: Majorana representation is not unique in
any sense.  There are infinitely many choices of the Dirac matrices
which satisfy \Eqs{clifford}{gamdag}.\footnote{In fact, there are
  other representations in which all four Dirac matrices are purely
  imaginary.  These can be obtained by any interchange of the matrices
  for $\gamma^1$, $\gamma^2$ and $\gamma^3$ that are given in
  \Eqn{majorep}, with the option of changing the overall sign of any
  number of them.}  An important theorem says that if there are two
choices of Dirac matrices, both satisfying \Eqs{clifford}{gamdag},
they will be related by a similarity transformation involving a
unitary matrix.  In other words, the general solution of
\Eqs{clifford}{gamdag} can be obtained from the Majorana
representation as
\begin{eqnarray}
\gamma^\mu = U \tilde \gamma^\mu U^\dagger 
\label{UgamUdag}
\end{eqnarray}
where $U$ is a unitary matrix.  If $\tilde\Psi$ is a solution of the
Dirac equation in the Majorana representation of the Dirac matrices, a
solution in this general representation will be given by
\begin{eqnarray}
\Psi = U \tilde \Psi \,,
\label{Upsi}
\end{eqnarray}
as can be checked easily from \Eqn{DiracEq}.

So, how will the Majorana condition, \Eqn{psi=psi*}, appear if we
choose to work with some other representation of the Dirac matrices
except the Majorana representation?  From \Eqs{psi=psi*}{Upsi}, we can
easily find that the condition would be
\begin{eqnarray}
U^\dagger \psi = \Big( U^\dagger \psi \Big)^\star
\end{eqnarray}
or
\begin{eqnarray}
\psi = UU^\top \psi^\star \,.
\label{UUpsi}
\end{eqnarray}
Note that since $U$ is unitary, the combination $UU^\top$ is also
unitary.  Instead of using $U$ directly, it is customary to use
another unitary matrix $C$ which is defined by
\begin{eqnarray}
UU^\top = \gamma_0 C \,,
\label{defC}
\end{eqnarray}
and create a compact notation for denoting the kind of combination
that the right hand side of \Eqn{UUpsi}:
\begin{eqnarray}
\hat\Psi \equiv \gamma_0 C \Psi^\star \,.
\label{defhat}
\end{eqnarray}
While this notation can be used for any fermion field, a Majorana
fermion field is defined through the condition
\begin{eqnarray}
\psi = \hat\psi \,.
\label{psi=psihat}
\end{eqnarray}
%

\subsection{Fourier expansion}
In the Majorana representation, the solution $\tilde\psi$ is real.  In
analogy with \Eqn{kg.phi}, we can write down its Fourier expansion:
\begin{eqnarray}
\tilde\psi(x) = \sum_s \int_p \Big( a_s(p) \tilde u_s(p) e^{-ip\cdot
  x} + a_s^\dagger (p)\tilde u_s^\star(p) e^{+ip\cdot x} \Big) \,.
\label{MajoFouri}
\end{eqnarray}
Notice that the solution involves some basis objects $\tilde u_s(p)$
and $\tilde u_s^\star(p)$, which are spinors.  Since the Dirac
matrices are $4\times4$, we will need four basis spinors.  Two basis
spinors $\tilde u_s(p)$ and their complex conjugates should be able to
do the job.  This is why there is a sum appearing in \Eqn{MajoFouri}:
the index $s$ takes two values, corresponding to the two independent
basis spinors.  It has to be said that the independent variables in
the spinors are the components of the spatial vector $\vec p$, since
the energy is related to the 3-momentum.  However, we will write
$u(p)$ and so on when no confusion arises.

Notice that the two terms of \Eqn{MajoFouri} are obviously conjugates
of each other, which is how the expression should give a real
$\tilde\psi$.  The question is, how will this Fourier expansion look
in an arbitrary representation for Dirac matrices?  Using \Eqn{Upsi},
we obtain
\begin{eqnarray}
\psi (x) =  \sum_s \int_p \Big( a_s(p) U \tilde u_s(p) e^{-ip\cdot x} +
a_s^\dagger (p) U \tilde u_s^\star(p) e^{+ip\cdot x} \Big) \,.
\label{psi}
\end{eqnarray}
Let us now define the basis spinors for the arbitrary representation
through the relation
\begin{eqnarray}
u_s(p) = U \tilde u_s(p) \,,
\end{eqnarray}
which mimics \Eqn{Upsi} for the field operator.  Obviously then,
\begin{eqnarray}
U \tilde u_s^\star(p) = U \Big( U^\dagger u_s(p) \Big)^\star = UU^\top
u_s^\star(p) \,.
\end{eqnarray}
We can therefore write \Eqn{psi} as
\begin{eqnarray}
\psi (x) = \sum_s \int_p \Big( a_s(p) u_s(p) e^{-ip\cdot x} +
a_s^\dagger (p) v_s(p) e^{+ip\cdot x} \Big) \,,
\label{psiv}
\end{eqnarray}
introducing the notation
\begin{eqnarray}
v_s(p) = \gamma_0 C u_s^\star(p) \,,
\label{u*=v}
\end{eqnarray}
where the matrix $C$ was defined in \Eqn{defC}.  Taking the complex
conjugate of both sides and multiplying by $UU^\top$, it is easy to
see that this definition also implies
\begin{eqnarray}
u_s(p) = \gamma_0 C v_s^\star(p) \,.
\label{v*=u}
\end{eqnarray}
In the Majorana representation, \Eqn{u*=v} and \Eqn{v*=u} means the
same thing, viz., the $u$ and the $v$ spinors are complex conjugates
of each other.

\subsection{Some properties of the matrix $C$}
The matrix $C$ has some interesting properties which we want to derive
now.  Using the definition of \Eqn{defC}, we obtain
\begin{eqnarray}
C^{-1} \gamma_\mu C &=& U^\star U^\dagger \gamma_0 \gamma_\mu \gamma_0
UU^\top 
\nonumber\\ &=& 
U^\star U^\dagger \gamma_\mu^\dagger UU^\top
\nonumber\\ &=& 
U^\star \Big( U^\dagger \gamma_\mu U \Big)^\dagger U^\top
\nonumber\\ &=& 
U^\star \tilde \gamma_\mu^\dagger U^\top 
\nonumber\\ &=& 
\Big( U \tilde \gamma_\mu^\star U^\dagger \Big)^\top \,,
\end{eqnarray}
where we have used \Eqs{gamdag}{UgamUdag}.  Finally now,
using \Eqn{gamstar}, we obtain
\begin{eqnarray}
C^{-1} \gamma_\mu C &=& - \Big( U \tilde \gamma_\mu U^\dagger
\Big)^\top = - \gamma_\mu^\top \,,
\label{CgamC}
\end{eqnarray}
which can also be taken as a definition for the matrix $C$.  In this
form, the definition does not refer to the Majorana representation at
all.  Combining this with \Eqn{defhat}, we now obtain a definition of
Majorana fermion that is independent of any representation.  No matter
which representation of Dirac matrices you are working with, you can
find the matrix $C$ in that representation through \Eqn{CgamC} and use
it to define $\hat\psi$ through \Eqn{defhat}.  A Majorana fermion
satisfies \Eqn{psi=psihat}, which is a generalized form of the
straight-forward reality condition of \Eqn{psi=psi*}.

The second interesting property of the matrix $C$ can be derived by
noting that, since $U$ is unitary,
\begin{eqnarray}
UU^\top U^* U^\dagger = 1 \,.
\end{eqnarray}
From \Eqn{defC}, it can be rewritten as
\begin{eqnarray}
\gamma_0 C (\gamma_0 C)^* = 1 \,.
\end{eqnarray}
Using \Eqs{gamdag0i}{CgamC}, this can be written as
\begin{eqnarray}
- C C^* = 1 \,,
\end{eqnarray}
or equivalently as
\begin{eqnarray}
C^* = \null -C^{-1} \,.
\end{eqnarray}
Using the unitarity of the matrix $C$, this relation can be cast into
the form
\begin{eqnarray}
C^\top = \null -C \,,
\label{Ctrans}
\end{eqnarray}
i.e., $C$ must be an antisymmetric matrix in any representation of the
Dirac matrices.

\subsection{Lorentz invariance of the reality condition}
Let us now go back to the reality condition of \Eqn{psi=psihat}.  The
condition would be physically meaningful only if it holds irrespective
of any reference frame, i.e., is Lorentz invariant.  We now show that
this is indeed the case.

Under infinitesimal Lorentz transformations which take the co-ordinate
of a spacetime point from $x^\mu$ to $x'^\mu = x^\mu + \omega^{\mu\nu}
x_\nu$, a fermion field transforms as follows:
\begin{eqnarray}
\Psi'(x') = \exp \left( -{i \over 4} \omega^{\mu\nu} \sigma_{\mu\nu}
\right) \Psi(x) \,,
\label{psi'}
\end{eqnarray}
where
\begin{eqnarray}
\sigma_{\mu\nu} = \frac i2 \Big[ \gamma_\mu , \gamma_\nu \Big] \,.
\label{sigma}
\end{eqnarray}
Taking the complex conjugate of \Eqn{psi'} and multiplying from the
left by $\gamma_0C$, we obtain
\begin{eqnarray}
\hat\Psi'(x') &=& \gamma_0 C \exp \left( + {i \over 4} \omega^{\mu\nu}
\sigma_{\mu\nu}^* \right) \Psi^*(x) \next
&=& \gamma_0 C \exp \left( + {i \over 4} \omega^{\mu\nu}
\sigma_{\mu\nu}^* \right) (\gamma_0 C)^{-1} \hat\Psi(x) \,.
\label{Psi'1}
\end{eqnarray}
This contains the complex conjugate of the sigma-matrices.  In order
to tackle them, let us note that \Eqs{gamdag}{CgamC} tell us that
\begin{eqnarray}
\gamma_\mu^* \equiv \Big( \gamma_\mu^\dagger \Big)^\top =
\gamma_0^\top \gamma_\mu^\top \gamma_0^\top = - (\gamma_0C)^{-1}
\gamma_\mu (\gamma_0C) 
\end{eqnarray}
This gives
\begin{eqnarray}
\gamma_0 C \sigma_{\mu\nu}^* (\gamma_0 C)^{-1} = - \sigma_{\mu\nu} \,. 
\end{eqnarray}
Using this, we can simplify \Eqn{Psi'1} and write
\begin{eqnarray}
\hat\Psi'(x') 
&=& \exp \left( - {i \over 4} \omega^{\mu\nu} 
\sigma_{\mu\nu} \right) \hat\Psi(x) \,.
\label{Psihat'}
\end{eqnarray}
Apart from the hats on the fermion field, this equation is exactly the
same as \Eqn{psi'}.  In other words, this equation tells us that
$\hat\Psi$, defined in \Eqn{defhat}, transforms exactly the same way
that the fermion field does under proper Lorentz transformations.  The
combination $\hat\Psi$ can therefore be called the Lorentz-covariant
conjugate, or LCC, of $\Psi$.

It is now obvious why a reality condition like \Eqn{psi=psihat} is
Lorentz invariant.  Both sides of this equation transforms the same
way under Lorentz transformations.  So, if the condition is true in
any one Lorentz frame, it would be true in all frames.

\subsection{Generalization of the reality condition}
One can also make the following observation on \Eqn{psi=psihat}.  It
is Lorentz covariant of course, but if we put an extra numerical
factor on one side of the equation, it will still be Lorentz
covariant.  Constants with modulus not equal to unity can be
disallowed from normalization arguments, but we can still have a
condition of the form
\begin{eqnarray}
\psi = e^{i\alpha} \, \hat\psi .
\label{genreal}
\end{eqnarray}
No doubt this will also define a Majorana field.  The plane wave
expansion of this field will contain the phase $\alpha$.  Instead of
\Eqn{psiv}, we should now write
\begin{eqnarray}
\psi (x) =  \sum_s \int_p \Big( a_s(p) u_s(p) e^{-ip\cdot x} +
e^{i\alpha} a_s^\dagger (p) v_s(p) e^{+ip\cdot x} \Big) \,.
\end{eqnarray}
However, it is easy to see that the phase $\alpha$ cannot be
physically relevant.  Rather than working with the field $\psi$
satisfying \Eqn{genreal}, we can $e^{-i\alpha/2}\psi$ as our field,
and then this field will satisfy \Eqn{psi=psihat}.  Nevertheless, the
freedom is sometimes useful in some manipulations.

\section{Left or right?}
This is one of the frequently asked questions (or FAQ's, an acronym
made popular by internet sites), or maybe a frequently answered
question (i.e., FAQ in a different sense) even when no one asks it.
The literature seems to be replete with statements where a Majorana
neutrino is called either a left-handed fermion or a right-handed one,
thus volunteering an answer for its handedness, without even anyone
asking for it.

There is, of course, nothing wrong in answering a question before it
is asked, if someone feels that it is anticipated, and that the answer
would be helpful for understanding the topic under discussion.  The
problem here is that the answer makes no sense, because the implicit
question makes no sense.  To explain this statement, we need to get
into the definition of ``handedness''.  For particles obeying the
Dirac equation, there are two possible definitions, and we discuss
both in turn.

\subsection{Helicity}
A definition of ``handedness'' that can be applied to any particle has
to do with the relative orientation of its momentum and angular
momentum.  The definition hinges on a property called ``helicity''.
For a particle with 3-momentum $\vec p$, helicity is defined as
\begin{eqnarray}
  \label{helicity}
  h_{\vec p} \equiv {2\vec J \cdot \vec p \over \mag p} \,,
\end{eqnarray}
where $\vec J$ denotes the angular momentum of the particle, and $\mag
p = |\vec p|$.  The orbital part of the angular momentum is
perpendicular to the direction of momentum, and therefore does not
contribute to helicity.  Helicity can therefore be described as twice
the value of the spin component of a particle along the direction of
its momentum.  The factor of 2 is inserted in the definition so that
the eigenvalues of this operator come out to be integral for any
particle.

For a fermion obeying the Dirac equation, we can write the helicity as
\begin{eqnarray}
  \label{hdirac}
  h_{\vec p} = {\vec\Sigma \cdot \vec p \over \mag p} \,,
\end{eqnarray}
where $\frac12\vec\Sigma$ denote the spin matrices, which satisfy the
same commutation relations as the general angular momentum operators.
These matrices are given by
\begin{eqnarray}
  \Sigma^i = \frac12 \varepsilon^{ijk} \sigma_{jk} \,,
  \label{Sigma}
\end{eqnarray}
where $\sigma_{jk}$ are the space-space components of the set of
matrices $\sigma_{\mu\nu}$ defined in \Eqn{sigma}.\footnote{We take
  the convention $\varepsilon^{ijk}=+1$.  In order to avoid any
  possible confusion, we use neither the antisymmetric tensor nor the
  components of $\vec\Sigma$ with lower indices.}  It can be easily
seen that the eigenvalues of $h$ are $\pm1$.  An eigenstate with
eigenvalue $-1$ is usually called ``left-handed'', whereas an
eigenstate with eigenvalue $+1$ is called ``right-handed''.  In what
follows, we will often use the terms ``left-helical'' and
``right-helical'' instead, in order to avoid confusion.

The interesting point is that $h$ commutes with the Dirac Hamiltonian,
a fact that can be checked with very little effort from \Eqn{DiracH},
using the anticommutation relation of the Dirac matrices.  For a free
Dirac particle, helicity is therefore conserved: it does not change
with time.

Helicity is also invariant under rotations, as the dot product in its
definition clearly implies.  In other words, if an observer works with
a spatial co-ordinate system that is rotated with respect to that of
another observer, both of them will infer the same value of helicity
of a given particle.

However, helicity is not invariant under boosts.  This can be easily
seen by considering a simple example.  Consider a fermion whose spin
and momentum are both in the same direction, which we call the
$x$-direction.  Its helicity will be $+1$ in this case.  Now consider
the same particle from the point of view of a different observer who
is moving also along the $x$-direction, faster than the particle with
respect to the original frame.  For this observer, the particle is
moving in the opposite direction, so the unit vector along the
particle momentum is in the negative $x$-direction.  The spin,
however, does not change, since \Eqn{Sigma} tells us that the
$x$-component of spin is really the $yz$-component of a rank-2
antisymmetric tensor, and components perpendicular to the frame
velocity remain unaffected in a Lorentz transformation.  The result is
that, in the frame of this new observer, the helicity of the same
fermion turns out to be $-1$.  And the lesson is this: a massive
fermion cannot be exclusively left-helical or right-helical.  Helicity
depends on the observer who is looking at it.

We want to point out that in expounding this lesson, we have used the
phrase ``massive fermion''.  One might wonder where the question of
mass came into the argument.  The answer is that, in our simple
example, the different value of helicity is obtained from the point of
view of an observer who moves faster than the particle in the original
frame.  For a massless particle, such a frame is impossible since the
massless particle would always move at the speed of light.  Hence, for
a massless particle, the value of helicity should be Lorentz
invariant.  This is an issue that will be discussed later.

\subsection{Chirality}
The Greek word ``chiros'' means ``hand''.  From this word, the word
``chirality'' has been coined.  Etymologically, ``chirality''
therefore means ``handedness''.  The meaning assigned to this
technical word is associated with the matrix $\gamma_5$ which
anticommutes with all Dirac matrices:
\begin{eqnarray}
  \label{gam5def}
[\gamma_5,\gamma_\mu ]_+ = 0 \qquad \forall \mu \,.
\end{eqnarray}
From the anticommutation relation between the $\gamma$-matrices, it
can be easily seen that the matrix
\begin{eqnarray}
\gamma_5 = i \gamma^0 \gamma^1 \gamma^2 \gamma^3
\label{gggg}
\end{eqnarray}
satisfies \Eqn{gam5def}.  An overall factor can be arbitrarily chosen
in this definition, and we have chosen it in such a way that the
matrix $\gamma_5$ has the properties
\begin{eqnarray}
  \label{gam5}
  \gamma_5^\dagger = \gamma_5 \,, \qquad (\gamma_5)^2 = 1 \,.
\end{eqnarray}
The last property guarantees that the matrices
\begin{eqnarray}
  \label{RL}
  L = \frac12 (1-\gamma_5) \,, \qquad  R = \frac12 (1+\gamma_5) \,, 
\end{eqnarray}
can act as projection matrices on fermion fields and spinors.  Such
projections are also often called ``left-handed'' and
``right-handed'', but we will use the terms ``left-chiral'' and
``right-chiral'' in order to avoid confusion.  So, given any object
$\Psi(x)$ that satisfies the Dirac equation, we can break it up into a
left-chiral and a right-chiral part,
\begin{eqnarray}
  \label{PsiL+R}
  \Psi = \Psi_L + \Psi_R \,,
\end{eqnarray}
where
\begin{eqnarray}
  \label{defPsiLR}
  \Psi_L = L\Psi \,, \qquad \Psi_R = R \Psi \,.
\end{eqnarray}
Alternatively, we can say that
\begin{eqnarray}
  \label{defPsiLRalt}
  L \Psi_L = \Psi_L \,, \qquad R \Psi_L = 0 \,, 
\end{eqnarray}
and a similar set of equations for $\Psi_R$.

It is important to note that if we consider a left-chiral solution of
the Dirac equation, it remains left-chiral under Lorentz
transformations.  This is guaranteed by the fact that
\begin{eqnarray}
  \label{gam5sig}
  [\gamma_5 , \sigma_{\mu\nu} ] = 0 \qquad \forall \mu,\nu \,,
\end{eqnarray}
which follows easily from \Eqn{gam5def}, in conjunction with
\Eqn{sigma}.  Thus, chiral projections can be made in a Lorentz
covariant way.  However, chirality is not conserved even for a free
particle, because $\gamma_5$ does not commute with the mass term in
the Dirac Hamiltonian.  This can be seen from the fact that the mass
term in the Dirac Hamiltonian contains only one Dirac matrix, and
therefore anticommutes, rather than commutes, with $\gamma_5$.  There
is no problem with the derivative term.  It contains two Dirac
matrices (one of them hidden in the definition of $\bar\Psi$) and
therefore commutes with $\gamma_5$.

In this way, chirality and helicity have somewhat opposite
characteristics: helicity is conserved for a free particle but is not
Lorentz invariant, whereas chirality is Lorentz invariant but not
conserved.  Therefore, none of these properties is appropriate for
characterizing a fermion that has mass.  If a particle is branded
left-helical, i.e., left-handed in the helicity sense, it will not
appear to be so to a suitably boosted observer.  If, at one time, a
particle is found left-chiral, i.e., left-handed in the chirality
sense, it will not remain so at another time.

\section{Weyl fermions}
It has been noted that the problem with assigning a frame-independent
helicity to a fermion disappears if the fermion is massless.  The
problem with a conserved value of $\gamma_5$ also disappears in this
limit, since $\gamma_5$ does indeed commute with the mass-independent
term in the Dirac Hamiltonian.  This shows that, without any
ambiguity, one can talk about a positive or negative helicity fermion
or of a left or right chiral fermion when one talks about massless
fermions.

\subsection{Irreducible fermion fields}
Indeed, it is very convenient to use such objects in any discussion
regarding fermions.  A general solution of the Dirac equation is not
an irreducible representation of the Lorentz group.  This is best seen
by the existence of the matrix $\gamma_5$ that commutes with all
generators of the representation, a fact that was summarized in
\Eqn{gam5sig}.  By Schur's lemma, no matrix other than the unit matrix
should have this property if the generators pertain to an irreducible
representation.  We have already seen that a left-chiral fermion field
retains its chirality under Lorentz transformations, implying that
such fields are irreducible.\footnote{Strictly speaking, this implies
  that left-chiral and right-chiral fields fall into different
  irreducible representations.  It does not preclude the possibility
  that either of these can be further reduced.  We show in
  Sec.\,\ref{ss:2c} that the chiral fields are indeed irreducible.}
So are right-chiral fields, of course.  It is known that the proper
Lorentz algebra is isomorphic to $\rm SU(2) \times SU(2)$, so that any
representation of the Lorentz algebra can be identified by its
transformation properties under each of the SU(2) factors.  In this
language, a left-chiral fermion would be a doublet under one of the
SU(2)'s and singlet under the other, a fact that is summarized by
denoting the representation as $(\frac12,0)$.  A right-chiral fermion
is a $(0,\frac12)$ representation.  Either of them is called a Weyl
fermion.  A general fermion field transforms like a reducible
representation $(\frac12,0)+(0,\frac12)$.  This tells us that a
general field can be described by two Weyl fields: one left-chiral and
one right-chiral.  This is the advantage of talking in terms of Weyl
fields: they can be seen as the building blocks for any fermion
field.

We could have said the same things in terms of helicity instead of
chirality, because there is a connection between the two in the
massless limit.  For massless particles, the Dirac equation for an
eigenstate of 3-momentum is given by
\begin{eqnarray}
  \label{masslesDEq}
  \Big( \gamma^0 \mag p - \vec \gamma \cdot \vec p \Big) w_p = 0 \,,
\end{eqnarray}
where $w_p$ can stand for $u(p)$ or $v(p)$.  This can be written as
\begin{eqnarray}
  \Big( 1 - \gamma^0  \vec \gamma \cdot {\vec p \over \mag p} \Big) w_p = 0
  \,.
\end{eqnarray}
Further, it can be shown that~\cite{repind}
\begin{eqnarray}
  \gamma^0 \gamma^i = \gamma_5 \Sigma^i \,,
\end{eqnarray}
so that we can write
\begin{eqnarray}
  \Big( 1 -  \gamma_5 {\vec \Sigma \cdot \vec p \over \mag p} \Big)
  w_p = 0 \,.
\end{eqnarray}
Multiplying throughout by $\gamma_5$ and using \Eqn{gam5}, we obtain
\begin{eqnarray}
  \gamma_5 w_p = {\vec \Sigma \cdot \vec p \over \mag p} w_p \,,
\label{c=h}
\end{eqnarray}
showing that helicity and chirality coincide for massless spinors.
Thus, we can talk about handedness of Weyl spinors without any
hesitation about the meaning of the term.

\subsection{Fourier expansion}\label{s:wffe}
Let us therefore talk about a left-handed Weyl fermion field.  In
order to keep a clear distinction with other kinds of fermions talked
about earlier, we will denote it by the symbol $\chi$.  It is
left-handed if it satisfies the relations
\begin{eqnarray}
L \chi = \chi \,, \qquad R \chi = 0 \,.
\label{lhfield}
\end{eqnarray}
The plane-wave expansion can now contain only left-chiral spinors $u_L
\equiv Lu$ and $v_L \equiv Lv$, and can be written as
\begin{eqnarray}
  \label{Weyl}
  \chi(x) = \int_p \Big( a(p) u_L(p) e^{-ip\cdot x} + \hat
  a^\dagger (p) v_L(p) e^{+ip\cdot x} \Big) \,.
\end{eqnarray}
Note that there is no sum over different solutions for $u$-type and
$v$-type spinors.  Since we have chosen one particular chirality,
there can be only one solution of each kind.  Because of this reason,
we have not put any index on the creation and annihilation operators
denoting spin projection or chirality.  

We now find out the helicities of the states that are produced from
the vacuum by $a^\dagger$ and $\hat a^\dagger$.  For this, we need to
invert \Eqn{Weyl}, and we need the explicit form of the momentum
integration implied in that equation.  Let us suppose that
\begin{eqnarray}
\int_p \equiv \int d^3p \; I_p \,,
\end{eqnarray}
where the factor $I_p$ depends only on the magnitude of the
3-momentum, or equivalently on the energy.  Let us also suppose that
the $u$-type and the $v$-type spinors have been normalized according
to the conditions
\begin{eqnarray}
& u_s^\dagger(\vec p)  u_{s'}(\vec p) = N_p \delta_{ss'} \,, \qquad 
v_s^\dagger(\vec p)  v_{s'}(\vec p) = N_p \delta_{ss'} \,, \next*
& u_s^\dagger(\vec p)  v_{s'}(-\vec p) = v_s^\dagger(\vec p)
u_{s'}(-\vec p) = 0 \,.
\label{uvnorm}
\end{eqnarray}
It is then easy to see that 
\begin{eqnarray}
a(k) &=& {1 \over (2\pi)^3 I_k N_k} \; \int d^3x \; e^{ik\cdot x}
u_L^\dagger(k) \chi(x) \,, \next
\hat a^\dagger(k) &=& 
{1 \over (2\pi)^3 I_k N_k} \; \int d^3x \; e^{-ik\cdot x}
v_L^\dagger(k) \chi(x) \,.
\label{a}
\end{eqnarray}
The first of these equations implies
\begin{eqnarray}
a^\dagger(k) &=& {1 \over (2\pi)^3 I_k N_k} \; \int d^3x \;
e^{-ik\cdot x} \chi^\dagger(x) u_L(k) \,.
\label{adag}
\end{eqnarray}

For any field $\Psi(x)$ satisfying the Dirac equation, the angular
momentum operator should satisfy a relation of the form
\begin{eqnarray}
\Big[ \Psi(x) , J_{\mu\nu} \Big] = \left( i (x_\mu \partial_\nu -
  x_\nu \partial_\mu) + \frac12 \sigma_{\mu\nu} \right) \Psi(x)
\end{eqnarray}
The purely spatial part involving derivatives on the right hand side
is the orbital angular momentum, and the second part the spin.  From
this, it follows that
\begin{eqnarray}
\Big[ \Psi(x) , h_{\vec p} \Big] = {\vec\Sigma \cdot \vec p \over \mag
p} \; \Psi(x) \,,
\label{psihcomm}
\end{eqnarray}
since the orbital part does not contribute, as remarked earlier.
Taking the hermitian conjugate of this equation, we obtain
\begin{eqnarray}
\Big[ \Psi^\dagger(x) , h_{\vec p} \Big] = - \Psi^\dagger(x) \;
{\vec\Sigma \cdot \vec p \over \mag p}  \,,
\label{psidaghcomm}
\end{eqnarray}
using the fact that $h_{\vec p}$ is hermitian because
\Eqs{gamdag}{Sigma} imply that the matrices $\vec\Sigma$ are
hermitian.  In particular, this equation is valid for the Weyl field
$\chi(x)$.  From \Eqn{adag}, we then obtain
\begin{eqnarray}
\Big[ a^\dagger(k) , h_{\vec k} \Big] &=& {1 \over (2\pi)^3 I_k
  N_k} \; \int d^3x \; e^{-ik\cdot x} \Big[ \chi^\dagger(x) ,
h_{\vec k} \Big] u_L(k) \next
&=& - \; {1 \over (2\pi)^3 I_k
  N_k} \; \int d^3x \; e^{-ik\cdot x} 
\chi^\dagger(x) {\vec\Sigma \cdot \vec k \over \mag k} u_L(k) \,.
\end{eqnarray}
Since we are dealing with a massless field here, we can use \Eqn{c=h},
and then the fact that $\gamma_5L = -L$, obtaining
\begin{eqnarray}
\Big[ a^\dagger(k) , h_{\vec k} \Big] 
&=& {1 \over (2\pi)^3 I_k
  N_k} \; \int d^3x \; e^{-ik\cdot x} 
\chi^\dagger(x) u_L(k) = a^\dagger(k) \,.
\end{eqnarray}
Applying both sides of this equation on the vacuum and noticing that
the vacuum state does not have any momentum, we obtain
\begin{eqnarray}
h_{\vec k} a^\dagger(k) \Big|0\Big> = - a^\dagger(k) \Big|0\Big> \,,
\end{eqnarray}
which shows that the state $a^\dagger(k)|0\rangle$ has helicity
$-1$.  Performing an exactly similar calculation starting with
\Eqn{psihcomm}, we can show that the state $\hat
a^\dagger(k)|0\rangle$ has helicity $+1$.  In other words, the Weyl
field operator annihilates a negative helicity particle and creates a
positive helicity antiparticle.

This could have been guessed from the CPT theorem.  Under very general
conditions, any field theory is CPT invariant.  CPT invariance means
that if we consider a process in which an initial state $A$ consisting
of some particles goes to some final state $B$, and another process
involving the CP-conjugate of the final state particles going into the
CP-conjugate of the initial state particles, the amplitude of the two
processes should be equal.  Therefore, a necessary condition for CPT
invariance is that if a particle exists in a theory, its CP conjugate
has to exist in the theory as well.  The CP-conjugate of a left-chiral
particle is a right-chiral antiparticle.  Therefore, if the
annihilation operator in a field operator annihilates a left-handed
particle, the creation operator must create a right-handed
antiparticle.  A left-chiral Weyl fermion field has only these two
states.  If we consider a right-chiral Weyl fermion field, that will
have a right-chiral particle and its CP-conjugate, a left-chiral
antiparticle.

\subsection{Majorana fermions from Weyl fermions}
We have said earlier that Weyl fermions, being irreducible
representations of the proper Lorentz group, can be used as building
blocks of any kind of fermion field.  We can now ask a specific
question: how can a Majorana fermion field be built out of Weyl
fields? 

A Majorana fermion has mass.  Therefore, it must have both left- and
right-chiral components.  It is therefore clear that we will need a
left-chiral Weyl fermion field as well as a right-chiral one in order
to obtain a Majorana field.  However, the arrangement with the two
chiralities must be such that the Majorana condition,
\Eqn{psi=psihat}, is satisfied.  So the question boils down to this:
how can one arrange to have two Weyl fields of two chiralities such
that they satisfy the Majorana condition?

Before we can answer this question, we have some groundwork to do.  A
left-chiral Weyl field satisfies the equation
\begin{eqnarray}
(1 + \gamma_5) \chi = 0 \,,
\label{defLsp}
\end{eqnarray}
as seen in \Eqn{lhfield}.  We take the complex conjugate of this
equation and multiply to the left by $\gamma_0C$, which gives
\begin{eqnarray}
\gamma_0C (1 + \gamma_5^*) \chi^* = 0 \,.
\label{gCom*}
\end{eqnarray}
Since $\gamma_5$ is hermitian, $\gamma_5^* = \gamma_5^\top$.
Using \Eqs{CgamC}{gggg}, we can show that
\begin{eqnarray}
C^{-1} \gamma_5 C  = \gamma_5^\top \,,
\label{Cgam5C}
\end{eqnarray}
or
\begin{eqnarray}
C\gamma_5^\top = \gamma_5 C \,.
\label{Cgam5T}
\end{eqnarray}
Thus, \Eqn{gCom*} can be written as
\begin{eqnarray}
\gamma_0 (1+\gamma_5) C\chi^* = 0 \,.
\end{eqnarray}
Since $\gamma_0$ anticommutes with $\gamma_5$, this is equivalent to
\begin{eqnarray}
(1-\gamma_5) \hat \chi = 0 \,,
\end{eqnarray}
using the definition of the Lorentz-covariant complex conjugation
given in \Eqn{defhat}.  This shows that if $\chi(x)$ is a
left-chiral Weyl field, $\hat\chi(x)$ is a right-chiral Weyl field.

The rest is obvious.  The complex conjugation operation is a toggle
operation, i.e., applying it twice is the same as not applying it
ever.  In other words,
\begin{eqnarray}
\hat{\left(\hat \Psi \right)} = \Psi 
\end{eqnarray}
for any kind of fermion field.  Thus, if we define a field by
\begin{eqnarray}
\psi(x) = \chi(x) + \hat\chi(x) \,,
\label{MajoWeyl}
\end{eqnarray}
it will obviously satisfy the reality condition of \Eqn{psi=psihat}
and will constitute a Majorana field.

The construction raises an interesting question.  A Weyl fermion is
massless whereas a Majorana fermion has mass.  How, by adding two Weyl
fermions with two opposite chiralities, we have also generated a mass? 

The point is that we have not really `generated' a mass: we have only
created an arrangement where mass can be allowed.  The mass term in
the Dirac Lagrangian is of the form $\bar\Psi\Psi$.  Using chiral
projections, it can be written as $\bar\Psi_L\Psi_R+\bar\Psi_R\Psi_L$.
There is no term like $\bar\Psi_L\Psi_L$ or $\bar\Psi_R\Psi_R$,
because these combinations vanish.  For a Weyl fermion field which has
a specific chirality, the mass term must therefore vanish.  In other
words, the mass term must contain two different chiralities: a Weyl
fermion is unable to meet this demand.

One might be tempted to argue that a term of the form $\chi^\top
C^{-1} \chi$ is Lorentz invariant, is quadratic in the fields, and
does not contain any derivatives.  Moreover, it can be constructed
with only one single chirality, and therefore should be considered as
a possible mass term for chiral fermion fields.  But this argument
does not work because this term is not hermitian.  One must add its
hermitian conjugate to the Lagrangian as well, and this can be written
as $\bar\chi\hat\chi$, or equivalently as $\hat\chi^\top C^{-1}
\hat\chi$.  Either way, it shows that we need the field $\hat\chi$ in
order to write down a mass term, and this $\hat\chi$ is a right-chiral
field, as shown before.  In summary, a massive fermion must have a
left-chiral as well as a right-chiral component.  A left-handed Weyl
fermion does not have a right-handed component, and hence cannot be
massive.  By adding a right-handed Weyl fermion $\hat\chi$ to the
left-handed $\chi$, we have fixed this shortcoming, and that is why a
Majorana fermion, given in \Eqn{MajoWeyl}, can have a mass.

\subsection{Dirac fermions from Weyl fermions}
Finally, we come to Dirac fermions.  These are also massive fermions,
and therefore require Weyl fermions of both helicities.  Also, these
are in general complex fermions, i.e., they do not satisfy any reality
condition like Majorana fermions do.  When we wrote \Eqn{MajoWeyl}, we
took the right-chiral Weyl field to be the LCC of the left-chiral
field, which is why the resulting field satisfied the reality
condition of \Eqn{psi=psihat}.  Instead, if we take two independent
left-chiral Weyl fields $\chi_1(x)$ and $\chi_2(x)$, and make the
combination
\begin{eqnarray}
\Psi(x) = \chi_1(x) + \hat\chi_2(x) \,,
\end{eqnarray}
this defines a Dirac field.

To summarize, a Dirac field is a completely unconstrained solution of
the Dirac equation.  Both Weyl and Majorana fields are simpler
solutions, with some kind of constraints imposed on the solution.  We
have seen that there are two types of conditions that can be imposed
in a Lorentz covariant manner on a solution of the Dirac equation.
One is a reality condition, imposition of which produces a Majorana
field.  The other is a chirality condition, imposition of which
produces a Weyl field.

We can ask whether we can impose both kinds of constraints at the same
time.  In other words, whether we can have a fermion field which is
both Majorana and Weyl.  It can be easily seen that it is not
possible.  The best way to see it is to use the Majorana
representation of the Dirac matrices.  In this representation, a
Majorana field is real.  On the other hand, a Weyl field 
satisfies either \Eqn{lhfield} or a similar equation obtained by
interchanging $R$ and $L$.  These two types of conditions can be
written in an alternative form,
\begin{eqnarray}
\gamma_5 \chi = \pm \chi.  
\label{gam5chi}
\end{eqnarray}
In Majorana representation of the Dirac matrices, $\gamma_5$ is purely
imaginary.  Therefore, \Eqn{gam5chi} cannot be satisfied by a real
field $\chi$, which shows that a Weyl field cannot be a Majorana field
at the same time.

\section{Two-component notation}\label{ss:2c}
In the Introduction, we mentioned that Weyl fermions can be
represented in a more compact notation, viz.\ as two-component
objects.  In this section, we discuss how this notation works, and how
much can be expressed with it.

\subsection{Weyl fermions}
To see why a 2-component notation would work for a Weyl fermion, let
us look at the Hamiltonian of \Eqn{DiracH}.  For massless particles,
$\gamma^0$ and $\gamma^i$ lose their separate identities: only the
combinations $\gamma^0\gamma^i$ appear in the Hamiltonian and
therefore can have physical consequences.  Let us use a shorthand to
denote these three matrices:
\begin{eqnarray}
\alpha^i = \gamma^0\gamma^i \,.
\label{alphai}
\end{eqnarray}
Using \Eqn{clifford}, it is easy to deduce the anticommutation
relations among the $\alpha^i$'s:
\begin{eqnarray}
\Big[ \alpha^i , \alpha^j \Big]_+ &=& 2 \delta^{ij} \,.
\end{eqnarray}
This set of relations can be satisfied by taking the $\alpha$'s to be
equal to the Pauli matrices.  In this representation, the solution of
the Dirac equation will be a 2-component object.

How are we going to define chirality of these objects, now that we
don't have the services of the matrix $\gamma_5$?  The answer is that
we take the help of the fact that chirality coincides with helicity
for massless spinors.  Helicity eigenstate spinors can be defined in
the 2-component notation by the equation
\begin{eqnarray}
{\vec \sigma \cdot \vec p \over \mag p} \, \xi_\pm(p) = \pm \xi_\pm
(p) \,, 
\label{heli2comp}
\end{eqnarray}
The solution $\xi_+$ has helicity eigenvalue $+1$, and is a
right-handed spinor.  The solution $\xi_-$ is left-handed.

The whole thing can also be viewed in terms of the $4\times4$
representation of the Dirac matrices.  Only, for this purpose, the
argument comes out clearly if we use a different representation of the
Dirac matrices, called the {\em chiral representation} or the {\em
  Weyl representation}.  In this representation, the Dirac matrices
are given by
\begin{eqnarray}
\Wrep\gamma^0 = \bmatrix{ 0 & 1 \cr 1 & 0 
} \quad &,& \quad 
\Wrep\gamma^i = \bmatrix{ 0 & \sigma^i \cr -\sigma^i & 0
} \,,
\label{chiralrep}
\end{eqnarray}
where we have put a crescent sign to indicate this representation.
From \Eqn{gggg}, it follows that in this representation,
\begin{eqnarray}
\Wrep\gamma_5 = \bmatrix{ -1 & 0 \cr 0 & 1 
} \,,
\label{g5chi}
\end{eqnarray}
i.e., is block diagonal.  This means
\begin{eqnarray}
\Wrep L = \bmatrix{ 1 & 0 \cr 0 & 0 
} \,, \qquad
\Wrep R = \bmatrix{ 0 & 0 \cr 0 & 1 
} \,.
\end{eqnarray}
Consider now the defining equation of a left-chiral spinor,
\Eqn{defLsp}.  Clearly, in this representation, the lower two
components of such a spinor must vanish.  For a right-chiral spinor,
the upper two components should vanish.  Thus, for a spinor with a
specific chirality, two of the four components are superfluous.  The
two-component notation described above essentially does away with the
vanishing components explicitly and deals only with the non-trivial
ones.

Surely, the Lagrangian as well as the equation of motion can be
written in the 2-component notation.  For this, we start from the
4-component notation, and note that in the chiral representation, the
$\alpha$-matrices are given by
\begin{eqnarray}
\alpha^i = \bmatrix{ -\sigma^i & 0 \cr 0 & \sigma^i 
} \,,
\label{alpha4X4}
\end{eqnarray}
using \Eqs{alphai}{chiralrep}.  Consider now a massless field $\Psi$. 
\Eqn{Lag} tells us that its  Lagrangian should be 
\begin{eqnarray}
\ms L = i \bar\Psi \gamma^\mu \partial_\mu \Psi =  i \Psi^\dagger
\Big( \partial_0 + \alpha^k \partial_k \Big) \Psi 
\end{eqnarray}
in any representation of the Dirac matrices.  We now introduce a
notation that we will use for arbitrary fermion fields:
\begin{eqnarray}
\Wrep\Psi = \bmatrix{\xi_t \cr \xi_b} \,,
\label{Psi2comp}
\end{eqnarray}
where $\xi_t$ and $\xi_b$ are 2-component fields.  They represent the
top two and the bottom two components of the fermion field in the
chiral representation.  Using \Eqn{alpha4X4}, we see that the
Lagrangian can be written down as
\begin{eqnarray}
\ms L = i \xi_t^\dagger \Big( \partial_0 - \sigma^k \partial_k \Big)
\xi_t + i \xi_b^\dagger \Big( \partial_0 + \sigma^k \partial_k \Big)
\xi_b \,. 
\label{tbLag}
\end{eqnarray}
If we are talking of a right-handed Weyl field, $\xi_t=0$.  The
Lagrangian for $\xi_b$ is often written in the more compact form
\begin{eqnarray}
\ms L = i 
\xi_b^\dagger \sigma^\mu \partial_\mu \xi_b \,,
\label{bLag}
\end{eqnarray}
where one defines the set of four $2\times2$ matrices
\begin{eqnarray}
\sigma^\mu \equiv (1, \vec\sigma) \,.
\label{sigmu}
\end{eqnarray}
For left-handed Weyl fields, the corresponding Lagrangian is
\begin{eqnarray}
\ms L = i 
\xi_t^\dagger \bar\sigma^\mu \partial_\mu \xi_t \,,
\label{tLag}
\end{eqnarray}
where
\begin{eqnarray}
\bar\sigma^\mu \equiv (1, -\vec\sigma) \,.
\label{barsigmu}
\end{eqnarray}
Note that this use of the bar has nothing to do with the definition
used for the fields, first introduced in \Eqn{Lag}.

The Dirac equation for Weyl fields can be written either from the
Lagrangians in \Eqs{bLag}{tLag} or starting from the massless Dirac
equation in the 4-component formalism.  Either way, one gets the
equations for the 2-component fields to be
\begin{eqnarray}
\sigma^\mu \partial_\mu \xi_t = 0 \,, \qquad 
\bar\sigma^\mu \partial_\mu \xi_b = 0 \,.
\end{eqnarray}
or more explicitly 
\begin{eqnarray}
\Big( \partial_0 - \sigma^i \partial_i \Big) \xi_t = 0 \,, \qquad 
\Big( \partial_0 + \sigma^i \partial_i \Big) \xi_b = 0 \,.
\end{eqnarray}

Lastly, let us write down the matrices $\sigma^{\mu\nu}$ in the
chiral representation.  It is straight forward to deduce, using
\Eqn{sigma}, the results
\begin{eqnarray}
\Wrep\sigma^{0k} = \bmatrix{ i\sigma^k & 0 \cr 0 &
    -i\sigma^k } \quad &,& \quad 
\Wrep \sigma^{ij} = \varepsilon^{ijk} \bmatrix{
    \sigma^k & 0 \cr 0 & \sigma^k } \,.
\label{sigChirep}
\end{eqnarray}
This shows that the generators are block diagonal, in two $2\times2$
blocks.  This is an explicit demonstration of the fact that the
$4\times4$ representation is reducible, a fact that we mentioned
earlier.  The $2\times2$ blocks are irreducible, since they contain
the Pauli matrices, which cannot be diagonalized simultaneously.

The expressions of $\Wrep \sigma^{ij}$ in \Eqn{sigChirep} explains
something that we have used but did not explain.  Note that in
\Eqn{hdirac} we used the matrices $\vec\Sigma$ to define helicity, but
then in \Eqn{heli2comp} we used the matrices $\vec\sigma$ in their
place.  This was done in anticipation of \Eqn{sigChirep}, which
implies that $\Wrep{\vec \Sigma}$ reduces to $\vec\sigma$ while
operating either on left-chiral fields which have only the two upper
components or on right-chiral fields which have only the two lower
components.

\subsection{Majorana fermions}
A Majorana fermion field, in the Majorana representation, has four
real components.  Can it also be represented in terms of a
2-component field if we allow for complex components?

To answer this question, let us look at the matrix $\Wrep U$ that
connects the chiral representation of the Dirac matrices to the
Majorana representation in the sense of \Eqn{UgamUdag}.  From the
explicit forms of the Dirac matrices in the two representations, it is
straight forward to show that
\begin{eqnarray}
\Wrep U = \frac12 \bmatrix{1+\sigma^2 & -i(1 - \sigma^2) \cr
  i(1-\sigma^2) & 1 + \sigma^2} \,.
\label{Uchi}
\end{eqnarray}

If a Majorana field is represented by the components
$\tilde\psi_1,\cdots,\tilde\psi_4$ in the Majorana representation, in
the chiral representation the field will be obtained by using
\Eqn{Upsi}, and the explicit form of $\Wrep U$ from \Eqn{Uchi}.  The
result is
\begin{eqnarray}
\Wrep\psi = \frac12 \pmatrix{
\vspace{2mm}
(\tilde\psi_1+\tilde\psi_4) - i(\tilde\psi_2+\tilde\psi_3) \cr 
\vspace{2mm}
(\tilde\psi_2-\tilde\psi_3) + i(\tilde\psi_1-\tilde\psi_4) \cr
\vspace{2mm}
-(\tilde\psi_2-\tilde\psi_3) + i(\tilde\psi_1-\tilde\psi_4) \cr 
(\tilde\psi_1+\tilde\psi_4) + i(\tilde\psi_2+\tilde\psi_3) 
}  \,.
\label{chpsi}
\end{eqnarray}
The components in the Majorana representation are real.  We notice
that, by taking real and imaginary parts of only the two upper
components of $\Wrep\psi$, we can obtain all the information that is
there in $\tilde\psi$.  For example, $\tilde\psi_1 =
\mbox{Re}\,\Wrep\psi_1 + \mbox{Im}\,\Wrep\psi_2$.  In a similar
manner, we can determine all components of $\tilde\psi$ explicitly in
terms of $\Wrep\psi_1$ and $\Wrep\psi_2$ only.  The two lower
components are not independent because $\Wrep\psi$ should satisfy the
Majorana condition.  In the chiral representation,
\begin{eqnarray}
\Wrep \gamma^0 \Wrep C = \bmatrix{0 & i\sigma^2 \cr -i\sigma^2 & 0 }
\,, 
\label{Cchirep}
\end{eqnarray}
which can be easily checked through \Eqs{defC}{Uchi}.  Thus, if we
use the right hand side of \Eqn{Psi2comp} to represent a Majorana
field, we obtain the relation between the top and bottom 2-components
in the form
\begin{eqnarray}
\xi_b = -i\sigma^2 \xi_t^* \,, \qquad \xi_t = i\sigma^2 \xi_b^* \,, 
\label{btreln}
\end{eqnarray}
from the Majorana condition, \Eqn{psi=psihat}.  These relations can be
explicitly checked in the expression of $\Wrep\psi$ given in
\Eqn{chpsi}.  Thus we can write a Majorana field in the chiral
representation in the form
\begin{eqnarray}
\Wrep\psi(x) = \bmatrix {\omega(x) \cr -i\sigma^2 \omega^*(x)} \,.
\label{defomega}
\end{eqnarray}
Clearly, everything about the Majorana field can be written down 
by using the upper two components only, which we have denoted by
$\omega(x)$.  This is also a 2-component representation like that used
for Weyl fields earlier, only we use a different symbol in order to
avoid confusion.

The Lagrangian of a Majorana field, in the 4-component notation, is
given by
\begin{eqnarray}
\ms L = \frac12 \Big( \bar\psi \, i\gamma^\mu \partial_\mu \psi - m
\bar\psi\psi \Big) \,.
\label{MajoLag}
\end{eqnarray}
The overall factor of $\frac12$ compared to the general Dirac
Lagrangian is usual for self-conjugate fields, introduced to ensure a
consistent normalization of the field operators in quantum field
theory.  Using the representations of the Dirac matrices given in
\Eqn{chiralrep}, one obtains
\begin{eqnarray}
\ms L = \frac i2 \Big( \omega^\dagger \bar\sigma^\mu \partial_\mu
\omega - m \omega^\top \sigma^2 \omega \Big) + \mbox{h.c.} \,,
\label{2compL}
\end{eqnarray}
where ``h.c.'' means hermitian conjugate.  The equation of motion that
follows from this Lagrangian is given by
\begin{eqnarray}
\bar\sigma^\mu \partial_\mu \omega + m \sigma^2 \omega^* = 0 \,.
\label{2compEoM}
\end{eqnarray}

Let us give some details of the derivation of \Eqn{2compL} which might
be illuminating.  Consider the term containing the derivative of
$\psi$ in \Eqn{MajoLag}.  We already know, from
\3Eqs{tbLag}{tLag}{bLag}, how this term will look like if written in
the 2-component notation.  For a Majorana field since the top and
bottom two components are related by \Eqn{btreln}, we obtain
\begin{eqnarray}
\bar\psi \, i\gamma^\mu \partial_\mu \psi = i 
\omega^\dagger \sigma^\mu \partial_\mu \omega + i 
\omega^\top \sigma^2 \bar\sigma^\mu \sigma^2 \partial_\mu \omega^* \,.
\label{psikinetic}
\end{eqnarray}
The first term on the right-hand side appears, as it is, in
\Eqn{2compL}.  As for the second term, let us write it in the generic
form $a^\top Xb$, where $a$ and $b$ are column matrices and $X$ is a
square matrix.  Now let us put down explicit subscripts for the matrix
elements, i.e., write $a^\top Xb = a_\alpha X_{\alpha\beta} b_\beta$.
Suppose now we want to write it with the component of $b$ in front.
We have to interchange the places of $a_\alpha$ and $b_\beta$, but it
should be remembered that these are components of fermion fields, and
therefore they anticommute.  Thus we can write
\begin{eqnarray}
a^\top Xb = a_\alpha X_{\alpha\beta} b_\beta = - b_\beta
X_{\alpha\beta} a_\alpha = - b^\top X^\top a \,.
\end{eqnarray}
Using this, and  the relation
\begin{eqnarray}
\Big( \sigma^2 \, \bar\sigma^\mu \, \sigma^2 \Big)^\top = \sigma^\mu 
\end{eqnarray}
which can be easily checked to be true, we find
\begin{eqnarray}
i\omega^\top \sigma^2 \bar\sigma^\mu \sigma^2 \partial_\mu\omega^* = 
-i(\partial_\mu\omega^\dagger) \sigma^\mu \omega \,,
\end{eqnarray}
which is the hermitian conjugate of the first term of the right-hand
side of \Eqn{psikinetic}.  The derivation of the mass term of
\Eqn{2compL} is very similar, and we do not give the details.

\subsection{Fourier expansion}
Since Weyl and Majorana fields can be written down in 2-component
representation, it should also be possible to write their Fourier
expansion in this representation.  For a left-handed Weyl field, this
expansion is
\begin{eqnarray}
  \label{Weyl2comp}
  \chi(x) = \int_p \Big( a(p) \xi_-(p) e^{-ip\cdot x} + \hat
  a^\dagger (p) \xi_-(p) e^{+ip\cdot x} \Big) \,.
\end{eqnarray}
For a right-handed Weyl field, basis spinors $\xi_+$ will appear in
the expansion.  The spinors $\xi_\pm$ were defined in
\Eqn{heli2comp}.  The explicit components can be easily found out from
the defining equation, viz., 
\begin{eqnarray}
\xi_- = \pmatrix{e^{-i\varphi} \sin {\vartheta \over 2}  \cr -\cos
  {\vartheta \over 2}} \,, \qquad  
\xi_+ = \pmatrix{\cos {\vartheta \over 2}  \cr e^{+i\varphi} \sin
  {\vartheta \over 2}} \,,
\end{eqnarray}
where the components of the 3-vector $\vec p$ are denoted by
\begin{eqnarray}
p_x &=& \mag p \sin\vartheta \cos\varphi \,, \next
p_y &=& \mag p \sin\vartheta \sin\varphi \,, \next
p_z &=& \mag p \cos\vartheta \,.
\end{eqnarray}
The overall phases of $\xi_\pm$ have been adjusted so that
\begin{eqnarray}
\xi_+ = -i\sigma^2 \xi_-^* \,, \qquad
\xi_- = i\sigma^2 \xi_+^* \,.
\label{xiconj}
\end{eqnarray}

For a Majorana field, both helicities should be present in the Fourier
expansion, since a massive particle cannot have a Lorentz invariant
value of helicity, as discussed earlier.  Hence we start by writing
the Fourier expansion in the form~\cite{Case:1957zza}
\begin{eqnarray}
  \omega(x) = \sum_{r=1,2} \int_p \Big( a_r(p) \zeta_r(p) e^{-ip\cdot x} + 
  a_r^\dagger (p) \eta_r(p) e^{+ip\cdot x} \Big) \,,
\label{Majo2Fouri}
\end{eqnarray}
mimicking the 4-component expression of \Eqn{MajoFouri}, using
2-component basis spinors $\zeta_r(p)$ and $\eta_r(p)$.  Some
conjugation relations exist between the 2-component $\zeta$- and
$\eta$ spinors, similar to the relations between the 4-component $u$-
and $v$-spinors, \Eqs{u*=v}{v*=u}.  To find these relations, let us
first note that
\begin{eqnarray}
\bar\sigma^\mu \partial_\mu e^{\pm ip\cdot x} = \pm \Big( E +
\vec\sigma \cdot \vec p \Big) e^{\pm ip\cdot x} \,,
\end{eqnarray}
so that, substituting \Eqn{Majo2Fouri} into \Eqn{2compEoM} and
equating the coefficients of $a_1$ and $a_2$ respectively, we obtain
the relations~\cite{Case:1957zza}
\begin{eqnarray}
\eta_r = {E - \vec\sigma \cdot \vec p \over m} \; i\sigma^2
\zeta_r^* \,.
\label{zeta2eta}
\end{eqnarray}
Equivalently, one can write
\begin{eqnarray}
\zeta_r = - {E - \vec\sigma \cdot \vec p \over m} \; i\sigma^2
\eta_r^* \,,
\label{eta2zeta}
\end{eqnarray}
using the identities
\begin{eqnarray}
\left( {E - \vec\sigma \cdot \vec p \over m} \right)^{-1} = {E +
      \vec\sigma \cdot \vec p \over m} \,,
\end{eqnarray}
and 
\begin{eqnarray}
\vec\sigma^* = - \sigma^2 \, \vec\sigma \, \sigma^2 \,.
\end{eqnarray}

Any linearly independent choice of the $\zeta_r$'s, along with the
corresponding $\eta_r$'s defined from \Eqn{zeta2eta}, will constitute
the Fourier expansion of a Majorana field in the 2-component notation.
It would be instructive to examine the nature of the Fourier modes by
making specific choices for the basis spinors.  For example, let us
take
\begin{eqnarray}
\zeta_1 = \xi_- \,, \qquad \eta_2 = \xi_- \,.
\label{2basischoice}
\end{eqnarray}
Using the definition of $\xi_-$ from \Eqn{heli2comp} and the
conjugation property from \Eqn{xiconj}, we then obtain
\begin{eqnarray}
\eta_1 = - {E - \mag p \over m} \; \xi_+ \,, 
\qquad
\zeta_2 = {E - \mag p \over m} \;  \xi_+ \,. 
\end{eqnarray}
Despite having a factor of $m$ in the denominator, these expressions
vanish rather than diverge in the limit of vanishing mass, because
\begin{eqnarray}
{E - \mag p \over m} = {m \over E + \mag p} \,.
\end{eqnarray}
Thus, the Fourier expansion of \Eqn{Majo2Fouri} can be written
explicitly as
\begin{eqnarray}
  \omega(x) &=& \int_p \Big( a_-(p) \xi_-(p) e^{-ip\cdot x} + {m \over E
    + \mag p} \; a_+(p) \xi_+(p) e^{-ip\cdot x} \next* 
&& \qquad   + a_+^\dagger (p) \xi_-(p) e^{+ip\cdot x} + {m \over E
    + \mag p} \;  a_-^\dagger (p) \xi_+(p) e^{-ip\cdot x} \Big) \,.
\label{explMajoFouri}
\end{eqnarray}
Note that the subscript on $a^\dagger$ is the opposite of the
subscript of $\xi$ that multiplies it.  This has to do with the fact
that the helicity of the state produced by this part of the Fourier
expansion is opposite to the helicity of the spinor that appears in
the term, as proved in \S\,\ref{s:wffe}.

A look at the Fourier expansion of \Eqn{explMajoFouri} shows something
quite interesting.  In the non-relativistic limit, $\mag p\approx 0$
and so $E \approx m$, so that $m/(E+\mag p) \approx 1$.  In the static
limit, this is an exact result.  For such values of momenta, the two
chiralities are produced and annihilated with the same amplitude.
However, in the ultra-relativistic limit, $m\ll E$, so that the terms
involving $\xi_+$ becoming vanishingly small and the Majorana field
behaves very much the same as a left-chiral Weyl field.  Had we made
the opposite choices for $\zeta_1$ and $\eta_2$ in \Eqn{2basischoice},
i.e., taken $\xi_+$ instead of $\xi_-$, the resulting Majorana field
would have behaved like a right-chiral Weyl field in the
ultra-relativistic limit.

We can try to construct the Fourier decomposition of the 4-component
representation of the Majorana field from that of $\omega(x)$, using
\Eqn{defomega}.  The task is simple and straight forward.  Take
$\omega(x)$ as in \Eqn{Majo2Fouri}, form $-i\sigma^2\omega^*(x)$, and
put one on top of the other.  The resulting expression will have the
form given in \Eqn{psiv}, with the 4-component spinors given by
\begin{eqnarray}
\Wrep u_r(p) = \bmatrix{\zeta_r(p) \cr -i\sigma^2 \eta_r^*(p)} \,,
\qquad  
\Wrep v_r(p) = \bmatrix{\eta_r(p) \cr -i\sigma^2 \zeta_r^*(p)} \,.
\end{eqnarray}
Using the form of the matrix $\gamma^0C$ in the chiral representation
that was given in \Eqn{Cchirep}, it is straight forward to check that
the $u$- and the $v$-spinors indeed satisfy the conjugation relations,
\Eqs{u*=v}{v*=u}.  Using the relations between the $\eta$ and the
$\zeta$ spinors, one can also write the $u$- and the $v$-spinors in
the form 
\begin{eqnarray}
\Wrep u_r(p) = \bmatrix{\zeta_r(p) \cr {E + \vec\sigma \cdot \vec p
    \over m} \zeta_r(p)} \,, \qquad  
\Wrep v_r(p) = \bmatrix{\eta_r(p) \cr - {E + \vec\sigma \cdot \vec p
    \over m} \eta_r(p)} \,. 
\end{eqnarray}
With the choice of the basis spinors given in \Eqn{2basischoice}, this
would read
\begin{eqnarray}
\Wrep u_-(p) = \bmatrix{\xi_- \cr {m \over E + \mag p} \; \xi_-}
\,, \qquad &&
\Wrep u_+(p) = \bmatrix{{m \over E + \mag p} \; \xi_+ \cr \xi_+}
\,, \next[4mm]
\Wrep v_+(p) = \bmatrix{- {m \over E + \mag p} \; \xi_+ \cr \xi_+}
\,, \qquad &&
\Wrep v_-(p) = \bmatrix{\xi_- \cr - {m \over E + \mag p} \; \xi_-}
\,.
\end{eqnarray}
In the limit of vanishing mass, these spinors approach the eigenstates
of the matrix $\gamma_5$ given in \Eqn{g5chi}.

Finally, there should be a few words of caution about the use of the
2-component representation.  First, a Dirac field does not have a
similar 2-component representation.  The point is that a Dirac field
has in general four independent complex components.  The number of
independent parameters is half as much for a Weyl field because of the
chirality condition, and also half as much for a Majorana field
because of the reality condition, which can be accommodated in two
complex components.  But there is not enough room for the components
of a Dirac field.

For Weyl and Majorana fields, even though the 2-component
representation is more compact, for practical purposes of performing
calculations, it is more convenient to use the 4-component
representation given earlier.  The reason is that there is hardly a
physical process in which all particles are Weyl fermions or Majorana
fermions.  All charged fermions are Dirac fermions.  If we have to use
the 4-component representation of the Dirac matrices to deal with
them, it is convenient to use the same for the other kinds of fermions
as well~\cite{Schechter:1981hw}.

\section{Charge conjugation and CP}
A field theory is called charge conjugation symmetric if its action
remains invariant after substituting all fields by their complex
conjugates (with a phase factor, if necessary).  For a scalar field
$\phi(x)$, the charge conjugation operation obviously implies
replacement by a phase factor times $\phi^\dagger(x)$.  For a fermion
field, naturally, the Lorentz covariant conjugation has to be
involved, because otherwise the resulting action will not even be
Lorentz invariant.  So, the operation of charge conjugation $\ms
C$ is given by
\begin{eqnarray}
\ms C \Psi(x) \ms C^{-1} = \eta_C \hat\Psi(x) \,,
\label{CPsiC}
\end{eqnarray}
where $\eta_C$ is a phase factor.  This kind of symmetries can have
particular importance for Majorana fields, for which $\psi$ and
$\hat\psi$ are equal.  We will address this point shortly.

Before that, it is useful to discuss what the charge conjugation
operation means for chiral projections of fermion fields.  We first
note that the operation of charge conjugation is unitary, which implies
that it will also have to be linear.  Linearity of an operator $\ms O$
implies that
\begin{eqnarray}
\ms O \Big( \alpha_1 \ket{a_1} + \alpha_2 \ket{a_2} \Big) = 
\alpha_1 \ms O \ket{a_1} + \alpha_2 \ms O \ket{a_2} \,,
\end{eqnarray}
where $\ket{a_1}$ and $\ket{a_2}$ are arbitrary states, whereas
$\alpha_1$ and $\alpha_2$ are arbitrary complex numbers.  In other
words, a linear operation does not affect the numerical
co-efficients.  The `states', in this context, means anything that the
operator $\ms O$ acts on.

Take $\Psi_L \equiv L\Psi$ now.  The chiral projection matrix $L$ was
defined in \Eqn{RL}, and is a numerical matrix, i.e., a matrix whose
elements are numbers, not fields.  The operation $\ms C$ acts on
fields, and we should then write
\begin{eqnarray}
\ms C \Psi_L \ms C^{-1} = \ms C L\Psi \ms C^{-1} = L \ms C \Psi \ms C^{-1} 
= \eta_C L\hat\Psi \equiv \eta_C \hat\Psi_L \,.
\label{CPsiLC}
\end{eqnarray}
The co-ordinate $x$ is unaffected by $\ms C$, so we have not written
it in this equation.  

To see an important feature of \Eqn{CPsiLC}, let us also find out what
is the LCC of $\Psi_L$.  Using the definition of \Eqn{defhat}, we
find,
\begin{eqnarray}
\hat{(\Psi_L)} \equiv \gamma_0 C (\Psi_L)^\star = \gamma_0 C L^\star
\Psi^\star \,.
\end{eqnarray}
We now use the fact that $\gamma_5$ is hermitian, so that $L^\star =
L^\top$, and then use \Eqn{Cgam5T} to write $CL^\top = LC$.  Then,
using the anticommutation of $\gamma_0$ and $\gamma_5$, we obtain
\begin{eqnarray}
\hat{(\Psi_L)}  = \gamma_0 L C \Psi^\star = R \gamma_0  C \Psi^\star 
= R \hat\Psi \equiv \hat\Psi_R \,.
\label{PsiLhat}
\end{eqnarray}
In passing, we should also note that the same kind of relation holds
between the spinors.  For example, \Eqn{v*=u} implies that 
\begin{eqnarray}
  u_L = L \gamma_0 C v^* = \gamma_0 R C v^* = \gamma_0 CR^\top v^* =
  \gamma_0 C v_R^* 
\end{eqnarray}
and similar relations of this sort.

Let us review what we have obtained in \Eqs{CPsiC}{PsiLhat}.  On an
unconstrained fermion field, we found that the charge conjugation
operation and the LCC operation work the same way, apart from a
possible phase factor.  But these two operations are not at all the
same thing on chiral projections of fermion fields.  Nevertheless,
because of the equivalence of these two operations on unconstrained
fields, the literature is replete with instances \cite{noref} where
the two have been confused for chiral fields as well.  The confusion
is enhanced by using the notation $\Psi^c$ for the LCC and the charge
conjugate of a field $\Psi$ interchangeably, or sometimes without any
clarification, even for a Weyl fermion.  Elaborate statements are even
made to the effect that charge conjugation changes chirality.  This
makes no sense whatsoever, and can be best seen with Weyl fields for
which chirality is the same as helicity.  Helicity involves spin and
momentum, none of which changes under charge conjugation.  Thus
helicity is unaffected by charge conjugation, and so must be
chirality.

To avoid such confusion, I think it is best not to use the notation
$\Psi^c$ at all, neither for charge conjugation nor for LCC.  That is
what I have been doing in this article: I have denoted LCC by a hat,
and used the elaborate notation $\ms C \Psi(x) \ms C^{-1}$ when I
wanted to denote the charge conjugate of a field $\Psi(x)$.

The situation is different if we consider not just charge conjugation
but the combined operation $CP$.  Since parity operation involves the
matrix $\gamma_0$, for unconstrained fermion fields we can write
\begin{eqnarray}
\ms {CP} \; \Psi(t, \vec x) \; (\ms {CP})^{-1} = \eta_{CP} \gamma_0
\hat\Psi(t, -\vec x) \,, 
\label{CPPsiCP}
\end{eqnarray}
This is also a linear operation.  So, on left-chiral fields, it
implies 
\begin{eqnarray}
\ms {CP} \; \Psi_L(t, \vec x) \; (\ms {CP})^{-1} &=& L \ms {CP} \;
\Psi(t, \vec x) \; (\ms {CP})^{-1} = \eta_{CP} L \gamma_0
\hat\Psi(t, -\vec x) \next*
&=& \eta_{CP} \gamma_0 R
\hat\Psi(t, -\vec x) = \eta_{CP} \gamma_0
\hat\Psi_R (t, -\vec x) \,, 
\label{CPPsiLCP}
\end{eqnarray}
In this sense, $\hat\Psi_R$ is the CP-conjugate or the CP-transform of
$\Psi_L$.  This is what is naively expected: charge conjugation
changes particles to antiparticles and vice-versa, whereas parity
changes helicity, so that the CP-conjugate of a left-handed particle
is a right-handed antiparticle, as is seen from \Eqn{CPPsiLCP}.

Let us now discuss how C and CP symmetries can be employed in theories
involving fermion fields.  A free Majorana fermion is an eigenstate of
charge conjugation symmetry, as is clear from \Eqs{psi=psihat}{CPsiC}.
If the interactions of a certain Majorana fermion are also invariant
under this symmetry, then we can say that the physical Majorana
particle is also an eigenstate of charge conjugation.  This might work
to a good approximation for supersymmetric partner of the photon.
However, for neutrinos, the interactions violate charge conjugation
symmetry badly, so it is useless to ask whether a Majorana neutrino
can be an eigenstate of charge conjugation \cite{Kayser:1983wm,
  Kayser:1984ge}. 

The case of CP need not be the same.  As far as we know, CP symmetry
is respected to a good accuracy in nature.  Thus, to the extent that
we can ignore CP violation, we can think of a Majorana neutrino as a
CP eigenstate, in the sense that
\begin{eqnarray}
\ms {CP} \; \psi(t, \vec x) \; (\ms {CP})^{-1} = \eta_{CP} \gamma_0
\psi(t, -\vec x) \,.
\label{CPpsiCP}
\end{eqnarray}
This may not look like a typical eigenvalue equation because of the
presence of the matrix $\gamma_0$ on the right hand side, but it can
be shown \cite{Kayser:1984ge, MohaPal} that the particle states with
vanishing 3-momentum are indeed eigenstates of CP in every sense of
the term.

With a Weyl neutrino, it is not even possible to construct a theory
that is invariant under charge conjugation.  The reason should be
obvious from our earlier discussion about the effect of this operation
on Weyl fields.  If we have a left-handed Weyl field, its charge
conjugate should also be a left-handed field, with opposite internal
quantum numbers.  Existence of such an object is not implied or
guaranteed by the existence of the left-handed field.  If one wants
charge conjugation symmetry, one will have to add this extra object in
the theory.  For CP symmetry, this problem does not exist.  As we
said, the CP conjugate of a left-handed Weyl particle is a
right-handed Weyl particle, which is also the LCC.  Whatever
Lagrangian one writes with a field can also be written in terms of its
LCC field, so the CP conjugate is in the theory anyway.  This is
guaranteed by CPT conservation, as has been argued earlier.

Once we take CP-violating effects into account, even \Eqn{CPpsiCP}
cannot be used to define a Majorana fermion field.  In such cases, one
can fall back on CPT \cite{Nieves:1981zt}.  However, we want to point
out that even if CPT is not conserved, Majorana fermions can still be
defined.  The definition does not depend on any of the discrete
symmetries: it is a property of the proper Lorentz group that a
covariant conjugation rule can be defined.  Indeed, we have defined a
Majorana field in Sec.\,\ref{s:DE}, long before we have started the 
discussion on the discrete symmetries in this section.  For scalar
fields, it is obvious that one can impose the condition
$\phi=\phi^\dagger$ without the assistance of any discrete symmetry.
For a vector field like that of the photon or the $Z$-boson, we merely
say that the field is real, without making any reference to any
discrete symmetry.  For fermion fields, imposing the condition
$\psi=\hat\psi$ does not require anything more.  The proper Lorentz
group is respected by strong, weak and electromagnetic interactions,
so if this condition is imposed on the free field, it will remain
valid for the interactive field.

\section{Feynman rules}
Let us now discuss the Feynman rules for these different kinds of
fermions.  In view of the comments about the inconveniences
encountered in using the 2-component formalism for Weyl and Majorana
neutrinos, we will discuss the rules only in the 4-component formalism
which are valid in arbitrary representations of the Dirac matrices.

\subsection{Internal lines}
For internal lines, the propagator has to be used.  The propagator is
the amplitude of propagating from one spacetime point $x$ to another
point $y$.  In the language of quantum field theory, it is seen as the
annihilation of a particle at $x$ and its creation at $y$, and
therefore depends on the field operators $\Psi(x)$ and $\bar\Psi(y)$.
Of course the same operators can annihilate an antiparticle at $y$ and
produce it at $x$.  Both possibilities are entailed, depending on
whether $x$ or $y$ has a larger value of the time coordinate, in the
propagator.  The result depends on $x-y$, and we can take the Fourier
transform of it and denote the Fourier component with momentum $p$ by
the notation $S_p(\Psi_a\bar\Psi_b)$, where $a,b$ denote the component
of the fermion field.  For a Dirac field, the propagator is derived in
every book of quantum field theory and is given by
\begin{eqnarray}
S_p(\Psi_a\bar\Psi_b) = {\Big( \gamma^\mu p_\mu + m \Big)_{ab} \over
  p^2 - m^2} \,, 
\label{prop}
\end{eqnarray}
where the mass multiplies an implied unit matrix, and the numerator is
the $a,b$ matrix element of the matrix sum.  For Weyl fermions, the
mass has to be set to zero: that's all.

For Majorana fermions, the same expression is obtained as well.
However, there are more combinations of operators that can create a
particle at $x$ and annihilate it at $y$.  The reason is that a
Majorana particle is the antiparticle of itself, so that the field
operator $\psi(x)$ contains both the annihilation and the creation
operator of this particle.  Thus $\psi(x)$ can create as well as
annihilate, and so can $\bar\psi(x)$.  So, a propagator can be formed
even out of the field operators $\psi(x)\psi(y)$.  To obtain the
expression of such a propagator, note that \Eqn{psi=psihat} implies
\begin{eqnarray}
\psi^\top = \bar\psi C \,,
\end{eqnarray}
or in terms of matrix elements, 
\begin{eqnarray}
\psi_b = \bar\psi_d C_{db} \,.
\end{eqnarray}
Therefore,
\begin{eqnarray}
S_p(\psi_a\psi_b) = S_p(\psi_a\bar\psi_d) C_{db} = {\Big( \big(
  \gamma^\mu p_\mu + m \big) C \Big)_{ab} \over p^2 - m^2} \,, 
\label{prop..}
\end{eqnarray}
using the expression that appears in \Eqn{prop}.  Similarly, there can
also be the propagator with two $\bar\psi$ operators, and its
expression can be similarly obtained:
\begin{eqnarray}
S_p(\bar\psi_a\bar\psi_b) = S_p(\psi_d\bar\psi_b) (C^{-1})_{da} =
{\Big( C^{-1} \big(
  \gamma^\mu p_\mu + m \big) \Big)_{ab} \over p^2 - m^2} \,.
\label{prop--}
\end{eqnarray}

There is an interesting property of the propagators given in
\Eqs{prop..}{prop--} that is worth noticing.  As said earlier,
the components of fermion field operators anticommute, i.e.,
$\psi_a(x)\psi_b(y) = - \psi_b(y)\psi_a(x)$.  So, if we interchange
$x$ and $y$, and also the indices $a$ and $b$, the propagator should
change sign.  The Fourier transform kernel is $\exp(-ip\cdot(x-y))$,
so interchanging $x$ and $y$ implies changing the sign of $p$ in the
Fourier transform.  Thus we should have
\begin{eqnarray}
S_p(\psi_a\psi_b) = - S_{-p}(\psi_b\psi_a) \,,
\end{eqnarray}
and a similar equation for the $\bar\psi\,\bar\psi$ propagator.  In
other words, the matrices appearing in the expression with even powers
of $p$ should be antisymmetric, and those with odd powers of $p$
should be symmetric.  The mass term is easily seen to satisfy this
property since we have already proved that $C$ is antisymmetric, and
so $C^{-1}$ must also be so.  In addition, one can use
\Eqs{CgamC}{Ctrans} to show that both $\gamma^\mu C$ and
$C^{-1}\gamma^\mu$ are symmetric matrices.

\begin{table}
\caption{Feynman rules for external fermion
  lines.  For Majorana fermions, the phase $\alpha$ has been defined
  in \Eqn{genreal}.}\label{t:external}
\begin{center}
\begin{tabular}{|c|c|c|c|c|}
\hline

\multirow{3}{7.5em}{Type of fermion} &
\multicolumn{4}{|c|}{Feynman rule for} \\ \cline{2-5} 
 &  \multicolumn{2}{|c}{incoming} &
\multicolumn{2}{|c|}{outgoing} \\  \cline{2-5}
 & with $\psi$ & with $\bar\psi$  
 & with $\psi$ & with $\bar\psi$ \\
\hline 
Dirac particle & $\sum_s u_s(\vec p)$ & 0 & 0 & $\sum_s \bar u_s(\vec
p)$ \\ 
Dirac antiparticle & 0 & $\sum_s \bar v_s(\vec p)$ & $\sum_s v_s(\vec
p)$ & 0 \\  
Majorana & $\sum_s u_s(\vec p)$ & $e^{-i\alpha} \sum_s \bar v_s(\vec p)$ &
$e^{i\alpha} \sum_s  v_s(\vec p)$ & $\sum_s u_s(\vec p)$ \\
LH Weyl particle & $u_L(\vec p)$ & 0 & 0 & $\bar v_R(\vec p)$ \\
antiparticle of LH Weyl & 0 & $\bar v_R(\vec p)$ & $v_L(\vec p)$ & 0 \\ 
\hline
\end{tabular}
\end{center}
\end{table}
\subsection{External lines}
As for the case of internal lines, we do not elaborate on the external
line Feynman rules for Dirac fermions, since they are covered in any
standard textbook.  The rules for Weyl fermions resemble the rules for
Dirac fermions, with the only difference that there is only one spinor
for each case, and this spinor has a well-defined chirality.

For Majorana fermions, however, the rules are different.  The reason
for the difference has already been discussed while talking about the
propagators: the field operator $\psi$ can either create or annihilate
a particle, and so can $\bar\psi$.  The various possibilities that
arise have been summarized, along with the rules for Dirac and Weyl
particles, in Table~\ref{t:external}.

\subsection{An example}
The Feynman rules show why we need to be careful in dealing with
Majorana fermions.  The field operator can do multiple task which a
Dirac field operator cannot.  As an illustrative example, consider an
interaction 
\begin{eqnarray}
\ms L_{\rm int} = g \Phi \bar\Psi F \Psi \,,
\end{eqnarray}
where $g$ is a coupling constant, $\Phi$ is a boson field, and $F$ is
some $4\times4$ numerical matrix sandwiched between the fermion field
operators.  For example, if $\Phi$ is a spinless particle, $F$ can be
the unit matrix or $\gamma_5$, or a linear combination of the two.

If the boson is massive enough, it can decay into a final state
containing two fermions.  If the fermions are Dirac particles, the
Feynman amplitude for the process will be
\begin{eqnarray}
\ms M = g \sum_{s_1,s_2} \bar u_{s_1}(p_1) F v_{s_2}(p_2) 
\end{eqnarray}
with an obvious notation about the spins and momenta of the
final-state particles.  Here, the operator $\Psi$ creates the
antiparticle in the final state, whereas the operator $\bar\Psi$
creates the particle.

If, on the other hand, a Majorana pair is produced in the final state,
the amplitude will be different.  The reason is that, now the operator
$\psi$ can create either of the two, and so can $\bar\psi$.  So we
should write
\begin{eqnarray}
\ms M = g \sum_{s_1,s_2} \Big( \bar u_{s_1}(p_1) F v_{s_2}(p_2) - 
 \bar u_{s_2}(p_2) F v_{s_1}(p_1) \Big) \,,
\label{amplM}
\end{eqnarray}
omitting an overall factor of $e^{i\alpha}$ as dictated by
Table\,\ref{t:external}, because it would disappear anyway when the
absolute square of the amplitude will be taken to calculate any
physical quantity.  Notice also the relative minus sign between the
two terms, which appears because of the anticommutation relation of
the fermion fields.

This expression can also be written in an alternative form by using
\Eqs{u*=v}{v*=u}.  We note that, using shorthand notations like $u_1
\equiv u_{s_1}(p_1)$, we can write
\begin{eqnarray}
\bar u_2 F v_1 = (\gamma_0 Cv_2^*)^\dagger \gamma_0 F \gamma_0 C u_1^* 
= v_2^\top C^{-1} F \gamma_0 C u_1^* \,.
\end{eqnarray}
But the whole thing is a number, so we might as
well write it as the transpose of the matrices involved.  Thus,
\begin{eqnarray}
\bar u_2 F v_1 = \Big( v_2^\top C^{-1} F \gamma_0 C u_1^* \Big)^\top 
= u_1^\dagger C \gamma_0^\top F^\top  C^{-1} v_2 
= - \bar u_1 CF^\top  C^{-1} v_2 \,,
\end{eqnarray}
using \Eqn{CgamC} on the way.  The amplitude of \Eqn{amplM} can now be
written in the form
\begin{eqnarray}
\ms M = g \sum_{s_1,s_2} \bar u_{s_1}(p_1) \Big( F +  CF^\top  C^{-1}
\Big) v_{s_2}(p_2) \,. 
\label{F-CFC}
\end{eqnarray}

We started by saying that Majorana fermions are simpler objects
compared to Dirac fermions.  There cannot be any argument about this
statement, at least between persons who would agree that real numbers
and simpler than complex numbers, or a real scalar field is simpler
than a complex scalar field.  Yet, now we see that the amplitudes
involving Majorana fermions can have more terms compared to a similar
amplitude involving Dirac fermions, so there is a price to pay for the
simplicity.

It should be noted that this price has nothing to do with the
fermionic nature of the fields.  This is true even for scalar fields,
for example.  Consider an interaction term
$(\phi_1^\dagger\phi_2)(\phi_2^\dagger\phi_1)$ that drives a
tree-level elastic scattering between two bosons $\phi_1$ and
$\phi_2$.  If the fields are complex, there is only one way the
creation and annihilation operators can work for this process, viz.,
$\phi_1$ can annihilate the 1-particle in the initial state and
$\phi_1^\dagger$ can create it in the final state, and similarly for
$\phi_2$.  However, if the fields are real, there are more cases to
consider because then $\phi_1$ is the same as $\phi_1^\dagger$, and
any of the two factors of $\phi_1$ in the interaction term can
annihilate the initial state particle as well as create the final
state particle.  For the scalar case, such possibilities would produce
an overall factor, because everything else is the same.  For fermion
fields, because of the matrix structure, the different terms are not
exactly the same, but they are related, as we can see in the example
of \Eqn{F-CFC}.

{\small
\paragraph*{Acknowledgements~:} 
Amol Dighe enthusiastically responded when I proposed to give a
pedagogical talk in the NuGoa conference that he was organizing in Goa
in April 2009.  When I could not finish all I wanted to say in the
designated one hour's time, he even arranged a special evening session
where I could present the rest.  Gustavo Branco, one of the people in
that audience, asked me to give a similar set of lectures in Lisbon
when I visited his group in November 2009.  And Jorge Rom\~ao insisted
that I write up the material.  They all, along with the people in the
audiences, made me feel that these talks can be useful to people.  I
thank them all.  After the initial version of this write-up was
submitted to the internet, I have enjoyed discussions with Kaushik
Balasubramanian and Rainer Plaga.}

\end{document}